\documentclass[letter]{aa} 

\usepackage{graphicx}
\usepackage{txfonts}
\usepackage{orcidlink}
\usepackage[switch]{lineno}
\begin{document}

   \title{Temporal and chromatic variation of polarized scattered light\\ in the outer disk of PDS 70
   \thanks{based on observations collected at the European Southern Observatory under ESO programs: 110.240P.001, 095.C-0404(A), 096.C-0333(A), 099.C-0891(A), 0102.C-0916(B), 60.A-9801(S), and 1100.C-0481(T).}}
   \author{J. Ma \orcidlink{0000-0003-3583-6652}
          \inst{\ref{inst1}, \ref{inst2}}
    \and C. Ginski \orcidlink{0000-0002-4438-1971} \inst{\ref{inst3}, \ref{inst4}, \ref{inst5}}
    \and R. Tazaki \orcidlink{0000-0003-1451-6836}\inst{\ref{inst2}, \ref{inst6}} 
    \and C. Dominik \orcidlink{0000-0002-3393-2459}\inst{\ref{inst5}}
    \and H.M. Schmid \orcidlink{0000-0002-7501-4015}\inst{\ref{inst1}}
    \and F. Ménard \orcidlink{0000-0002-1637-7393}\inst{\ref{inst2}} 
    } 
    
   \institute{Institute for Particle Physics and Astrophysics, 
          ETH Zurich, Wolfgang Pauli Strasse 17, CH-8093 Zurich, Switzerland\\
              \email{jie.ma@univ-grenoble-alpes.fr}
              \label{inst1}
    \and{Univ. Grenoble Alpes, CNRS, IPAG, F-38000 Grenoble, France}\label{inst2} 
    \and {School of Natural Sciences, Center for Astronomy, University of Galway, Galway, H91 CF50, Ireland}\label{inst3} 
    \and {Leiden Observatory, Leiden University, PO Box 9513, 2300 RA, Leiden, The Netherlands}\label{inst4} 
    \and {Anton Pannekoek Institute for Astronomy, University of Amsterdam, Science Park 904, 1098XH Amsterdam, The Netherlands}\label{inst5} 
    \and {Astronomical Institute, Graduate School of Science, Tohoku University, 6-3 Aramaki, Aoba-ku, Sendai 980-8578, Japan} \label{inst6}
    }
   \date{Received 22 July 2024/ Accepted 25 October 2024}
 
  \abstract
   {PDS 70 stands out as the only system hosting a protoplanetary disk and two confirmed planets undergoing formation. It is a unique target for characterizing the dust in this type of disk.}
   {We aim to accurately measure the reflected polarized intensity and quantify the variability and asymmetry for PDS~70 across multiple epochs and wavelengths in the optical and near-infrared. We present new high-contrast polarimetric differential imaging observations of PDS~70, with the $N\_R$ filter on SPHERE/ZIMPOL. }
   {We combined the new observation with archival data of the VLT/SPHERE instrument, spanning five wavelengths ($N\_R$, $VBB$, $J$, $H$, and $Ks$) over seven epochs and eight years. For each observational epoch, we corrected the smearing effect due to finite instrument resolution, measured the azimuthal brightness profiles, and derived the intrinsic disk-integrated polarized reflectivity and the intrinsic brightness contrasts. }
   {
   With our homogeneous analysis of all available optical and near-infrared data sets of the disk around PDS 70, we find significant temporal variability of the integrated polarized reflectivity as well as the azimuthal brightness profile. This indicates variable shadowing on the outer disk by inner disk structures beyond the resolution limit of current imaging instruments. Despite these variabilities, we observe a systematic wavelength-dependent contrast between the near side and the far side of the inclined disk. These results underline the importance of considering both the shadowing effect from the inner disk and the surface geometry of the observed reflecting disk in the analysis and interpretation of observational data. }
   {}


   \maketitle

%

\section{Introduction}
\begin{figure*}[h]
    \centering
    \includegraphics[width=0.95\textwidth]{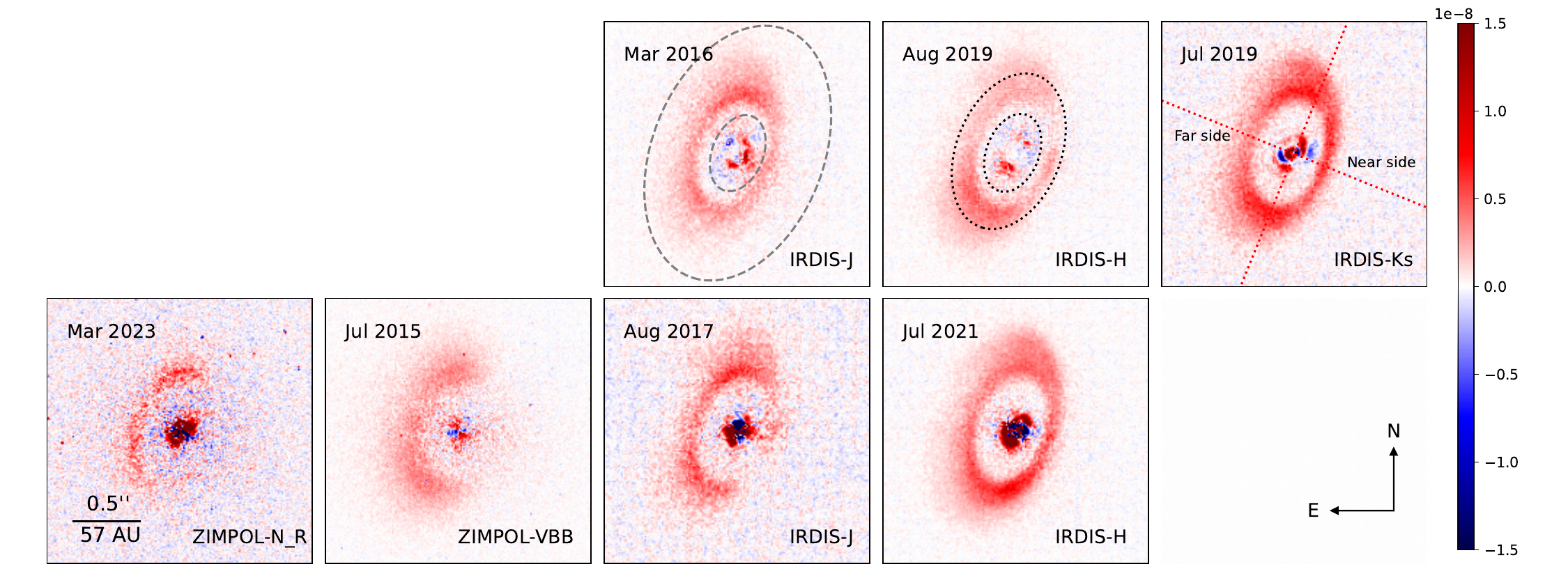}
    \vspace{-0.1cm}
    \caption{Mean polarized intensity images $Q_{\varphi}(x,y)$ for $N\_R$, $VBB$, $J$, $H$, and $Ks$ band. All images are normalized to the stellar intensity $I_{\star}$ and divided by their pixel areas ($3.6\times 3.6 \text{mas}^2$ for ZIMPOL and $12.25\times 12.25 \text{mas}^2$ for IRDIS) to provide the surface brightness contrast in $\text{mas}^{-2}$, as indicated by the color bar. The first and the second rows present coronagraphic and non-coronagraphic observations respectively. The shared color bar is indicated on the right. The epochs are indicated on the top left in each panel. The grey dashed line in the J-band panel indicates the integrated region for disk signals. The black dotted line in the $H$ panel indicates the deprojected region used to calculate the azimuthal profile. Up is to the north and left is to the east.}
    \vspace{-0.2cm}
    \label{fig: observations}
\end{figure*}

The study of dust properties in protoplanetary disks is crucial for understanding planet formation \citep{Birnstiel2024, Ginski2023, Andrews2018}. Multi-band polarimetric imaging offers a powerful tool to probe these properties, as it reveals the scattering behavior of dust grains in the disk's surface layers. Scattered light properties, such as disk reflectivity and colors \citep{Krist2005, Debes2013, Mulders2013, Tazaki2019, Hunziker2021, Ma2023, Ma2024}, as well as scattering phase functions \citep{Ginski2023, Tazaki2023} have been used to constrain the dust properties in the disk surface region. However, our ability to understand the dust properties via measurements from multi-wavelength polarised and/or total intensity observations often relies on the premise of stable, uniform illumination of the outer disk over different observation epochs. Variability in scattered light, particularly when it is due to shadowing effects, can significantly impact our observations and interpretation of dust properties. This requires accurately calibrated polarized reflectivities from multi-epoch imaging polarimetry. These measurements have been  scarcely available for protoplanetary disks until now.

PDS 70, a photometrically variable young K7 T-Tauri star of 5.4 Myr located 113.4 pc away in the Upper Centaurus-Lupus subgroup of the Sco-Cen star-forming region \citep{Muller2018, Pecaut2016, Gaia2018}, serves as a key object in this context. The system's well-defined inner and outer disks, separated by a clear gap, make it an ideal target for studying the influence of variable illumination on dust properties. The presence of two planets within the gap \citep{Keppler2018, Haffert2019, Benisty2021} further adds to its significant role as a laboratory for planet formation.
The spectral energy distribution (SED) fitting initially indicated the existence of these disks \citep{Metchev2004, Riaud2006}, with direct imaging in scattered and polarized light later resolving the outer disk and the large gap. The surface brightness of the ring-shaped outer disk peaks at 54 au, with an inclination of $i=49.7^\circ$ and a position angle of $68.6^\circ$ \citep{Hashimoto2012, Keppler2018}. The inner disk, although it remains unresolved, is estimated to have a radius of less than 18 au \citep{Keppler2018, Benisty2021}. Comparison between the dust continuum emission maps and reflected images reveals a similar morphology with dust size segregation \citep{Long2018, Portilla2022}.

Variability is seen in the PDS~70 system across various wavelengths and observing techniques.
The optical photometry shows both short-term variations in days and long-term on a yearly scale. The day-scale variation evolves between three-day periodic and stochastic "dipper-like" behavior, attributed to aperiodic occultation by micron-sized dust superposed on periodic stellar rotation \citep{Thanathibodee2020, Gaidos2024}. Infrared(IR) photometry dominated by inner disk emission also shows short and long-term variations, together with a one-year disappearance of emission at 3.4$\mu m$ and 4.6$\mu m$ in WISE data\citep{Gaidos2024}. In addition, strong mid-IR variation of 25\% is revealed when comparing JWST and \textit{Spitzer} SED with a 15-year interval \citep{Perotti2023, Jang2024}, and a comparable level of near-IR variability is seen in comparing JWST and SpeX \citep{Christiaens2024}, pointing to a dynamic inner disk geometry due to variable accretion. 
The scenario is supported by variable line emissions across multi-epoch \citep{Thanathibodee2020, Campbell-White2023, Gaidos2024}.
Moreover, multi-epoch ALMA dust continuum emission also suggests variable inner disk structure \citep{Casassus2022}.
The changing inner disk casts variable shadows on the outer disk, seen as azimuthally variable surface brightness in the scattered light and polarized light \citep{Juillard2022, Christiaens2024}. 

In this letter, we combine multi-epoch polarimetric imaging observations of PDS 70, revealing time variability in polarized reflectivity and azimuthal brightness profiles across 0.6 to 2.2 $\mu$m and exploring the potential chromatic dependence of the near-to-far side brightness contrast. Our work underscores the need for the careful interpretation of the measurements of scattered light.
\vspace{-0.25cm}
\section{Observations and data reduction}
\label{sect: observation}
For the first time in the narrow band $N\_R$ filter ($\lambda=645.9~nm$, $\Delta\lambda=56.7~nm$), PDS~70 was observed on March 1, 2023. This observation, using the SPHERE instrument on UT3 \citep{Beuzit2019}, is part of the optical scattered light survey (ID: 10.240P.001, PI: C. Ginski), aimed at characterizing the dust properties in young planet-forming disks. 
The observing sequence is designed to switch between long-exposures and short-exposures on target, combined with star-hopping to the reference star for long-exposures \citep{Wahhaj2021}. In this study, we selected all 11 polarimetric cycles of long-exposures, totalling up to 47~min of integration on target. 
The observation was performed in instrument-stabilized (P1) and the slow read-out mode \citep{Schmid2018}. 

We retrieved the archival data for PDS~70 observed with the ZIMPOL and IRDIS subsystem on the SPHERE instruments in polarimetric differential imaging mode. The observing condition and the data selections are summarized in Table~\ref{tab:obs-info} and detailed in Appendix~\ref{app: data-selection}. To summarize, we collected seven epochs of observations spanning seven years, with five color filters ($N\_R$, $VBB$, $J$, $H$, and $Ks$) from 0.6 to 2.2 $\mu$m. For simplicity, we labeled them by their band filter name, adding the epoch when two observations exist for the same band. Each epoch of observation contains full polarimetric cycles and non-saturated total intensity frames. For non-saturated polarimetric observations NR, VBB, and J17, the non-saturated total intensity is obtained simultaneously.  
For coronagraphic polarimetric observations J16 and Ks, two separate flux frames were taken before and after the polarimetric cycles. For coronagraphic H19, only one flux frame was taken after all polarimetric cycles. For the saturated polarimetric observation H21, eight pairs of non-saturated $Q^{+}$ and $Q^{-}$ frames are taken after the full polarimetric cycles. 

The ZIMPOL and IRDIS data are reduced with their dedicated pipelines sz-software \citep{Schmid2018} and IRDIS Data reduction for Accurate Polarimetry \citep[IRDAP,][]{VanHolstein2020}, respectively. 
On top of the processed data for each cycle, we conducted an extra step to minimize the integrated $U_{\varphi}$-signal within the disk region. This step should correct the telescope polarization and intrinsic stellar polarization for ZIMPOL and correct the stellar polarization for IRDIS. The disk region is defined as an elliptical aperture with $r =[0.3\arcsec, 1.0\arcsec]$, with the inclination $i=49.7^\circ$ and position angle PA$=68.6^\circ$ adopted from \citet{Hashimoto2012}, shown as the gray dashed lines in J16 panel of Fig.~\ref{fig: observations}. The estimated and subtracted polarization are summarized in Table~\ref{tab:intrinsic}.
As the last step, we calculate the azimuthal polarization parameters $Q_{\varphi}$ and $U_{\varphi}$ from the reduced Stocks polarizations $Q$ and $U$:
\vspace{-0.15cm}
\begin{align}
    Q_{\varphi} & = -Q\cos(2\varphi) - U\sin(2\varphi), \\
    U_{\varphi} & = Q\sin(2\varphi) - U\cos(2\varphi),
    \label{eq:uphi}
    \vspace{-0.1cm}
\end{align}
where $\varphi$ is the position angle to the central star from north over east. The mean $Q_{\varphi}(x,y)$ for each epoch are plotted in Fig.~\ref{fig: observations} and the corresponding $U_{\varphi} (x, y)$ are plotted in Fig.~\ref{fig: uphi}.
\vspace{-0.1cm}

\section{Results}
\label{sect: analysis}
Figure~\ref{fig: observations} shows temporal and chromatic variation of disk scattered light around PDS 70. Starting with the Ks image, which has the appearance of an archetypal transition disk, the disk near side is brighter than the far side, while it is nearly mirror symmetric to the minor axis. The near to far side brightness contrast is usually attributed to preferential forward scattering by grains large compared to the observing wavelength \citep[e.g.,][]{Duchene2004, Mulders2013}. However, this typical appearance is absent in other datasets. In the H bands (H19 and H21), the disk is not mirror symmetric with its southern ansa being brighter. Additionally, comparing H19 and H21, the disk flux relative to the stellar flux varies significantly, hinting at temporal variations in the system. More strikingly, at optical wavelengths (NR and VBB), the far side flux dominates over the near side. Consequently, the common picture of the bright side being the near side of the disk does not hold in this case, perhaps because the disk surface on the near side is at least partly hidden, since the illuminated inner disk wall is too steep to be visible \citep{Ma2022}. In the following sections, we will further quantify and discuss these temporal and chromatic variations. 
\vspace{-0.2cm}
\subsection{Integrated polarized reflectivity}
\begin{figure}
    \centering
    \includegraphics[width=0.45\textwidth]{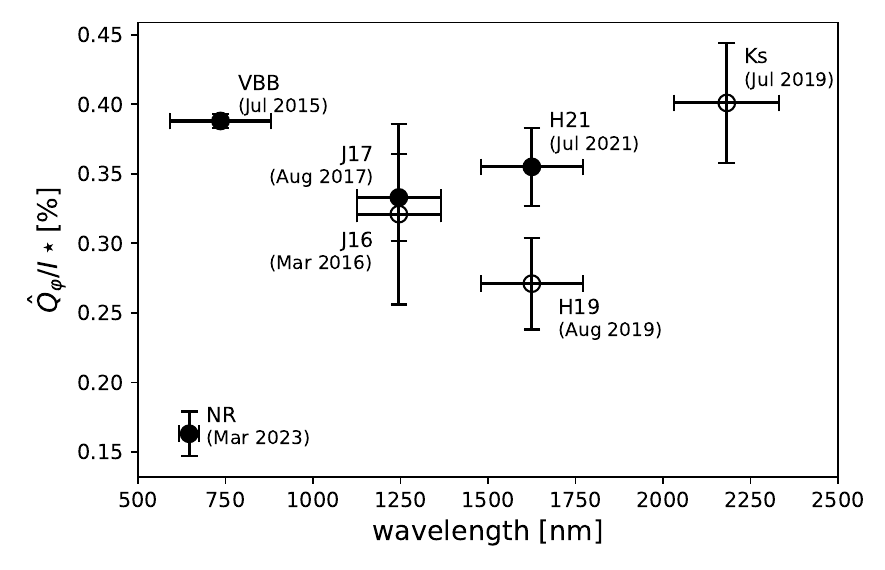}
    \vspace{-0.2cm}
    \caption{Corrected integrated polarized reflectivity $\hat{Q}_{\varphi}/I_{\star}$ as a function of wavelengths. The solid and hollow markers represent the values derived from non-coronagraphic and coronagraphic observations, respectively. The horizontal errorbars indicate the width of the band filters. The observing epochs are labeled beside each scatter. }
    \vspace{-0.25cm}
    \label{fig: reflectivity}
\end{figure}

\begin{figure}
    \centering
    \includegraphics[width=0.45\textwidth]{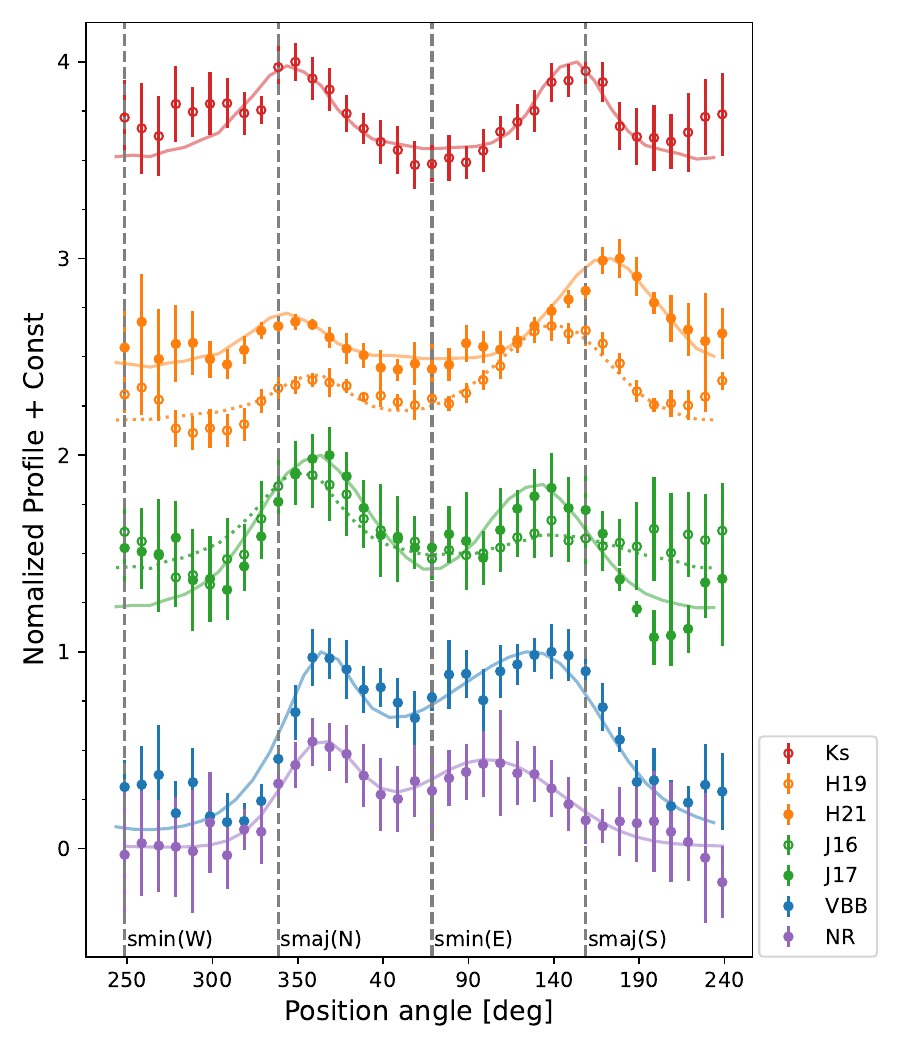}
    \vspace{-0.2cm}
    \caption{Azimuthal profile of the $Q_{\varphi}(\varphi)$ images averaged over an annulus of 35-70 au. The colored dots are calculated from the observations, with the uncertainties being the standard deviation over all selected cycles. The overplotted curves are calculated from the convolved fitted model. The vertical dashed lines indicate the position angles of the minor axis, smin(W) and smin(E), in the west and east directions, and the major axis, smaj(N) and smaj(S), in the north and south directions. }
    \vspace{-0.3cm}
    \label{fig: variation}
\end{figure}

As the first step, we integrated the total intensity $I_{\star} = \int I(x, y)$ within a 3$\arcsec$-diameter round aperture, to monitor the short-term stellar flux variability and calibrate the polarized disk flux. No significant temporal change in flux is observed for a given observing run lasting typically one hour. The integrated polarized flux $Q_{\varphi}$ is measured from the polarized intensity images $Q_{\varphi}(x,y)$ cycle by cycle within the elliptical aperture shown in Fig.~\ref{fig: observations} J16 panel. Table~\ref{tab:aperture-count} summarizes the integrated system flux $I_{\star}$, disk polarized flux $Q_{\varphi}$, and their ratio and the detailed derivation is available in Appendix~\ref{app:photometric-calib}.  

For accurate photo-polarimetric measurements, PSF smearing effects must be considered. The PSF smearing is crucial because it causes a blurring of the intrinsic disk signal and introduces polarimetric cancellation between regions with positive and negative Q and U signals \citep{schmid2006, Tschudi2021}. We followed the correction method described in \cite{Ma2024} to calculate the intrinsic disk-integrated polarized flux. Figure~\ref{fig: reflectivity} shows the corrected disk-integrated polarized reflectivity $\hat{Q}_{\varphi}/I_{\star}$ as a function of wavelength and the values are summarized in Table~\ref{tab:aperture-count}. Our results are consistent with the $\hat{Q}_{\varphi}/I_{\star}$ measurements of VBB, J17, H19, Ks data in \cite{Ma2024} considering we used a smaller integration region. 

At given wavelength, the $\hat{Q}_{\varphi}/I_{\star}$ measurements of J16 and J17 are consistent. In contrast, $\hat{Q}_{\varphi}/I_{\star}$ is about $30\%$ higher in H21 than in H19, and $140\%$ higher in VBB than in NR, not explained by the smearing. This inconsistency is also not caused by the coronagraph, as we would expect similar behavior in the J band. The likely explanation for this discrepancy is a structural change in the inner disk, changing the overall illumination on the outer disk, as discussed in Sect. \ref{section: dust}. 
\vspace{-0.2cm}

\subsection{Azimuthal profiles of polarized light}
\label{section: azimuthal}
For all seven observations, we derived the azimuthal profiles by averaging $Q_{\varphi}(x,y)$ over an annulus with a deprojected radius of 35-70 au in $10^\circ$ bins. The deprojected radius is calculated using diskmap \citep{Stolker2016}, assuming a power-law function of the disk surface $h(r)=0.041$au$\cdot (r/$1au$)^{1.25}$\citep{Ginski2023}. The annulus is shown in Fig.~\ref{fig: observations} H19 panel. The azimuthal profiles are plotted in Fig.~\ref{fig: variation}, starting from the near side PA=$246^\circ$. The profiles are normalized to the peak value at given wavelengths and shifted for clarity. The NR profile is normalized to the peak value at VBB without offset. The J-(J16, J17), H-(H19, H21), and Ks-band profiles are offset by 1, 2, and 3, respectively. The J profiles use the J17 peak for normalization, and H profiles use the H21 peak. The vertical dashed lines indicate the minor and major axis. Our $Q_{\varphi}(\varphi)$ profiles for VBB and J17 agree with the analysis of \citet{Keppler2018} across all position angles within the error margin for VBB and most angles for J17 except the near side. The comparisons are detailed in Appendix~\ref{app: polarimetric-calib}. 


Comparing the two azimuthal profiles at $J$ band, J16 and J17 marginally agree at most azimuthal angles within the errorbars. Except the J17 lacks the flux at PA$\sim200^\circ$, likely a local shadow, which leads to the integrated flux difference. H21 agrees with H19 only near PA$\sim130^\circ$ and $250^\circ$, while for all other azimuthal directions, H21 is brighter than H19, with the surface brightness maximas being approximately 80\% and 50\% brighter on the N and S sides, respectively. Such an overall brightness change also exists between NR from 2023 and VBB from 2015. The surface brightness maximas in VBB are about 120\% and 80\% brighter on the N and S sides than in NR. Furthermore, when comparing the azimuthal profiles across all wavelengths, we observe a tentative trend that the position of the surface brightness maxima shifts closer to the major axis for longer wavelengths, though this trend is partly mixed with time variability.

The PSF smearing effects also affect the azimuthal variance $Q_{\varphi}(\varphi)$ and need to be corrected.  
We describe the correction in Appendix~\ref{app: corr-azi} with an analytical disk model. The azimuthal profiles calculated from the convolved best-fitting model are overplotted in Fig.~\ref{fig: variation}. To conclude, the overall azimuthal profile shape is preserved after the convolution and the PSF smearing effect is not substantial, especially when comparing the contrast between position angles symmetric to the minor axis. 
We introduce $\hat{\alpha}_{X-Y}$ to quantify the brightness contrast between two azimuthal directions $X$ and $Y$, 
\vspace{-0.12cm}
\begin{equation}
    \hat{\alpha}_{X-Y} = \dfrac{\hat{Q}_{\varphi}(X)-\hat{Q}_{\varphi}(Y)}{\hat{Q}_{\varphi}(X)+\hat{Q}_{\varphi}(Y)} \, ,
\end{equation}
with $\hat{Q}_{\varphi}(X)$ being the averaged corrected polarized flux at $X$ direction in a $20^\circ$ wedge with deprojected radius 35-70 au. 

\begin{figure}
    \centering
    \includegraphics[width=0.45\textwidth]{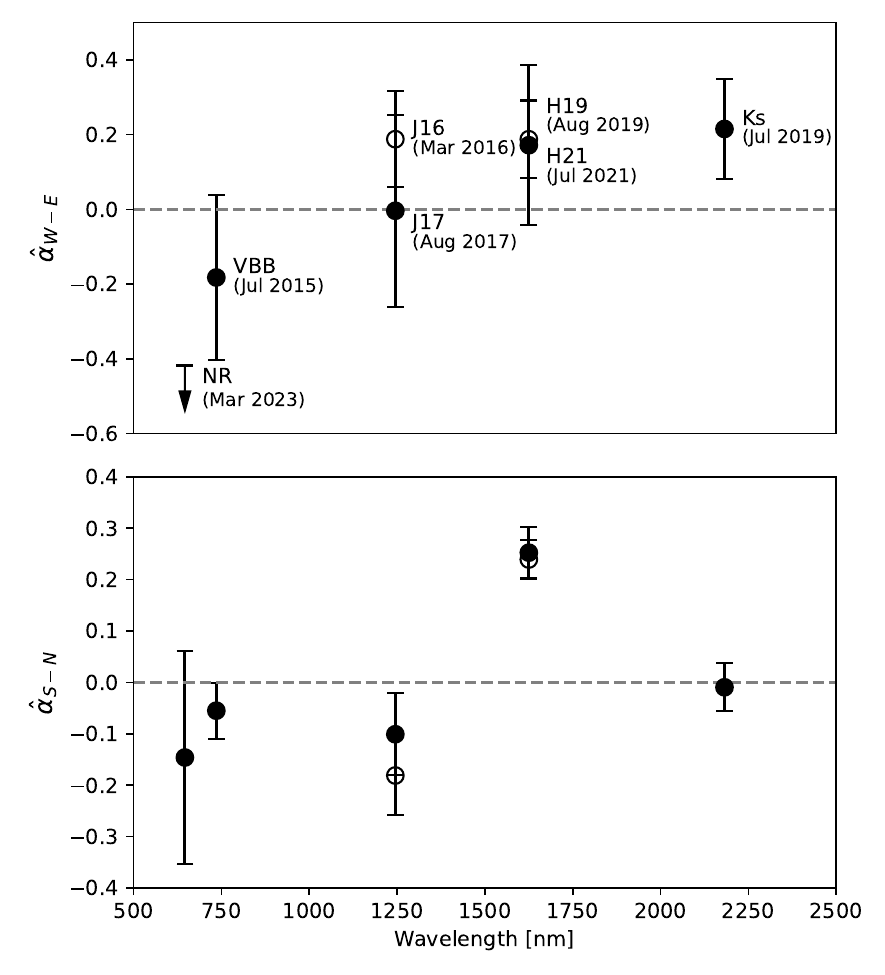}
    \vspace{-0.2cm}
    \caption{Brightness contrast of PDS~70. Upper: Brightness contrast between the W (near side) and the E (far side) along the minor axis. Lower: Brightness contrast between the brightness maximas on the S and the N close to the major axis. The errorbar on wavelength is omitted.}
    \vspace{-0.25cm}
    \label{fig: contrast}
\end{figure}

First, we measured the contrast between the west (W, near side, PA=$248.6^\circ$) and east (E, far side, PA=$68.6^\circ$) directions, denoted as $\hat{\alpha}_{W-E}$. 
The results are plotted in Fig.~\ref{fig: contrast} upper panel and the values are given in Table~\ref{tab: contrast}. This contrast tentatively increases with wavelength, exhibiting a faint near side in the visible, which becomes equally bright near the $J$ band, and brighter than the far side at the $H$ and $Ks$ band. 
We also calculated $\alpha_{W-E}$ without smearing effect correction for comparison with \citet{Keppler2018}. 
Our VBB $\alpha_{W-E} = -0.45\pm0.17$ aligns closely with \citet{Keppler2018} measurement of $-0.51$. However, we report for J17 $\alpha_{W-E} = 0.11\pm 0.13$, suggesting a slightly brighter W-side, whereas \citet{Keppler2018} reported $-0.17$. This small discrepancy is probably due to different normalization methods and uncertainty definitions. Despite this, the tentative trend of increasing near-to-far contrast with wavelength still holds. This well-defined effect can be caused by the wavelength dependence of the dust scattering properties or disk scattering geometry.

The second contrast was derived between the two brightest regions near the major axis N and S (PA = $\varphi_1$ and $\varphi_2$ are epoch-dependent, values are given in Table~\ref{tab: analytical}), denoted as $\hat{\alpha}_{S-N}$. The results are plotted in Fig.~\ref{fig: contrast} lower panel. This contrast shows no systematic trend with wavelength.
Ideally, a geometrically symmetric disk expects $\hat{\alpha}_{S-N}=0$ across all wavelengths. While equal brightness is seen for VBB and K, the W side is brighter for NR, J16, J17, and the E side is brighter for H19 and H21. 
Moreover, J16 and J17 have similar contrasts of $\hat{\alpha}_{S-N}=-0.18\pm 0.08$ and $-0.10\pm0.08$, respectively. And H19 and H21 have the same level of opposite contrast $\hat{\alpha}_{S-N}=0.24\pm 0.04$ and $0.25\pm0.05$ respectively. Therefore, despite the opposite brighter side comparing $J$ and $H$ bands, $\hat{\alpha}_{S-N}$ remains consistent at each given wavelength, without systematic trend over a time scale of a few years. 
\vspace{-0.2cm}

\section{Discussion}
\label{sect: discussion}
\subsection{Time variability}
\label{section: shadow}
Our multi-epoch polarimetric data reveal notable integrated polarized reflectivity variability and azimuthal brightness variability in the polarized scattered light from the PDS~70 outer disk. This finding is based on calibrated imaging polarimetry taking the PSF smearing effects carefully into account. This variability can be linked with the recently reported light curve variability by \citet{Gaidos2024}, which evolves between periodic and stochastic behavior, hinting at transitions between stable and unstable accretion regimes \citep{Gaidos2024}. 

In the stable accretion regime, the matter flows along the magnetic field as funnel streams that are lifted above the equatorial plane. This configuration creates broad and quasi-periodic shadows on the outer disk, as the elevated streams intercept a significant portion of the stellar light before it reaches the outer regions. Consequently, the outer disk appears fainter in scattered light due to the reduced illumination. While the periodic shadows would result in predictable patterns in the azimuthal brightness variability, our observations with the shortest intervals of 6 months, are not sensitive to the quasi-periodic shadowing of 3 days \citep{Thanathibodee2020}. 

On the other hand, during the unstable accretion regime, the system is dominated by Rayleigh-Taylor instabilities that drive matter through equatorial tongues \citep{Romanova2012}. Since these tongues are closer to the disk plane, they intercept less stellar light than the elevated funnels and more light reaches the outer disk, resulting in a brighter outer disk. The chaotic accretion and the thin tongue structures cause random and narrow shadows on the outer disk. This equatorial accretion alone is unlikely to obscure the star enough to explain the observed variability. Instead, a magnetized wind, as suggested by \citet{Gaidos2024}, may also contribute to the dimming and variability.

The switching is not only supported by the optical light curve, but also by significant IR variations observed over different time intervals. For instance, near-IR measurements have shown a variation of 25\% when comparing data taken years apart, indicating dynamic changes in the inner disk due to variable accretion \citep{Christiaens2024}. Additionally, mid-IR variability comparisons between JWST and \textit{Spitzer} also support this dynamic inner disk geometry \citep{Perotti2023}. However, mapping our observations to the variability is not possible due to the temporal gaps and the lack of high-cadence photometry during our observations. 

In the meantime, stellar flux variability also contributes to the observed integrated flux variation thought not directly recognizable in our observation. Repeated stellar flux measurements taken during our single observing epochs with a duration of about one hour remain stable at $\sigma(I_\star)/I_\star \approx 1\%$ and exhibit no systematic flux changes. This uncertainty falls short of resolving the periodic variation typically at 0.5\% per hour. Although we can exclude the occurrence of deep dimming events ($\Delta$g$>0.3^{m}$ in 1 day) during our observation, shallow dimming may still be present. 
\citet{Gaidos2024} also reported long-term behavior in the optical, showing an upper envelope of brightness oscillating on a yearly timescale by a few tenths of a magnitude, which could add variability.

Additionally, part of the azimuthal variance can be attributed to data reduction effects. For example, \citet{Ma2023} demonstrated that for RX~J1604, applying different intrinsic polarization subtraction can shift the broad shadowing features but does not contribute the integrated flux. Indeed, comparisons of our profiles with that of \citet{Keppler2018} reveal slight discrepancies (see Appendix~\ref{app: polarimetric-calib}). However, the general agreement across most position angles suggests that data reduction effects are minor for PDS~70 and the azimuthal brightness variability is primarily influenced by shadowing effects. Lastly, the impact of multiple scattering on the large-scale azimuthal dependence is also minimal (see the $U_{\varphi}$ images in Appendix~\ref{app: polarimetric-calib}). 

The time variability complicates the disk color measurement. Considering the higher fluxes as the unshadowed value at visible and $H$ band, 
PDS~70 shows a gray color from 0.6 to 2.2 $\mu$m. 
However, using NR data would indicate a red color, and H19 data would introduce an abrupt flux drop.  
The variability also complicates the extraction of scattering phase functions \citep{Ginski2023}. Time-dependent changes in the phase functions could lead to incorrect interpretation of dust grain size, composition, or distribution if not properly accounted for. Therefore, the measurements of reflectivity colors and scattering phase functions must be interpreted with caution for PDS 70 and other systems showing significant brightness variation for the reflected light from the disk.

To explore the time variability and structural changes in the inner disk more thoroughly, continuous photometric monitoring in both visible and IR wavelengths, paired with simultaneous high-contrast imaging and/or interferometry, will be essential. These approaches will allow us to determine the accretion regime, identify shadowing effects, and capture the structure of the inner disk, providing deeper insights into the mechanisms driving the observed variability.
\subsection{Wavelength dependence of brightness contrast}
\label{section: dust}
Despite variations in integrated flux and brightness profile, the brightness contrasts -- both between near and far and between brightness maxima at given wavelengths -- appear relatively stable over time based on the available observations. Here, we discuss some possibilities causing the observed tentative chromatic dependence from optical to near-IR. 

The first possible cause for the wavelength dependence of the near-to-far side asymmetry is the polarized intensity scattering phase function of dust grains in the disk surface. Polarized intensity is a product of the degree of polarization and total intensity; thus, the polarized intensity data alone cannot inform us of which specific factor is creating this wavelength dependence. Assuming the degree of polarization is wavelength-independent, grains have to be more efficient backscatterers at optical compared to near-IR. For example, by assuming the Heyney-Greenstein phase function and a plane-parallel radiative transfer model developed in \cite{Ma2022}, we estimated the asymmetry parameter of anisotropic scattering \citep{Bohren1983} to be around $g\approx -0.25$ in the optical to $g\approx 0.1$ in the Ks band (see Appendix \ref{app: flat-disk} for details). Backward-scattering particles is also suggested by \citet{Wahhaj2024} from PDS~70's $YJH$ total intensity images. Such a peculiar backward scattering behavior is not typically expected for small dielectric particles \citep[e.g., Rayleigh scattering; ][]{Bohren1983} or large fluffy dust aggregates composed of sub-micron monomers \citep{Volten2007, Tazaki2016}. One possible explanation is the presence of small metallic dust grains, which exhibit strong backward scattering \citep{vandeHulst1957, Jackson1975}. However, this would predict a similar degree of backscattering at longer wavelengths, which is not observed, making small metallic particles unlikely. Another intriguing possibility is the presence of large compact aggregates in the disk surface, which might also account for the enhanced backward-scattering behavior. \citet{Min2016} demonstrated that compact dust aggregates of 4 $\mu$m in size could exhibit a negative $g$ value if the observer only has access to intermediate scattering angles. In their simulations, a negative $g$ value does not appear for aggregates smaller than 2 $\mu$m (see Figs. 4 and 5). This increase in backward scattering is thought to result from the diffusive reflection on the rough dust surfaces, such as the lunar phase \citep{Mukai1982, Min2010}. Therefore, both large size and some degree of compactness of aggregates seem crucial for this phenomenon. However, quantitative conditions for the dust properties remain unclear and require future studies. If this explanation accounts for the observed negative $g$ value, the surface region of PDS 70 must be dominated by large compact grains or aggregates at least a few microns in size. On the other hand, the degree of polarization may change drastically with wavelength \citep{Tazaki2022}. A detailed investigation of such dust properties is beyond the scope of this paper, but observations of the degree of polarization are crucial to resolving parameter degeneracy in polarized intensity.

The second possibility is obscuration of the near-side scattered light by the inner edge of the outer disk. The scattering surface is expected to be at a progressively higher altitude for shorter wavelengths due to increased opacity, particularly notable for small grains, which can eventually cause self-obscuration of the near-side scattered light \citep{Ma2022}. Based on the plane-parallel radiative transfer model, we found that the wall slope needs to be $>90^\circ -i=40^\circ$ (see Appendix \ref{app: flat-disk}). Notably, this steep slope is consistent with the vertical gas structure constraints from recent CO isotopologue observations, supporting the feasibility of this kind of geometry\citep{Law2024}.

The third possibility is non-uniform outer disk illumination due to a misaligned inner disk. 
Misaligned disks have been suggested in various systems based on scattered light observations, 
\citep[e.g.,][]{Marino2015,Stolker2016b, Benisty2017, Pinilla2018, Muro2020}. PDS70 hosts an inner disk \citep{Keppler2019}, which potentially casts a shadow onto the outer disk. This shadowing effect is likely more efficient at shorter wavelengths (optical) due to higher opacity, while being partially transparent in the near-IR. However, in this case, the inner disk's tilt has to be fine-tuned so that the shadow covers exactly the near side of the outer disk. The stellar light passing through the atmospheric layer of the inner disk before reaching the outer disk could impose an additional polarization signal. The current models of PDS 70 typically assume an aligned or even absent inner disk and often overlook time variations in scattered light. These simplifications could lead to an incomplete understanding of the disk's geometry and the scattering properties of dust. Accounting for a misaligned inner disk and the time variability would be beneficial for developing more realistic models and interpretations of observed scattered light.

Stability is also observed in the brightness contrast between two maxima, $\hat{\alpha}_{S-N}$. 
Consistent with our measurements, PDS~70's HiCIAO PDI data from 2012 reported $\alpha_{S-N} = 0.13\pm0.06$ for H-band \citep{Hashimoto2015}. Meanwhile, SMA and ALMA observations of dust continuum emission indicate an azimuthal brightness enhancement on the N side at 0.87mm \citep{Long2018, Keppler2019} and 1.3mm \citep{Hashimoto2015}. 
If the asymmetric $\alpha_{S-N}(\lambda)$ is a time-independent feature, \citet{Hashimoto2015} proposed a long-lived dust clump induced by accreting planets to explain the opposite asymmetry between H-band scattered light and ALMA dust emission. However, the consistency of J band scattered light asymmetry with ALMA contradicts the explanation. 
These tentative time-independent features need further observations to confirm, especially simultaneous multi-wavelength observations to mitigate the impact of temporal variability and properly interpret the wavelength dependence in scattered light observations. Simultaneous total intensity observations will also help distinguish between shadowing effects and intrinsic dust property variations in the faint regions.

\vspace{-0.5cm}
\section{Conclusions}
\label{sect: conclusion}
We present the first polarized reflected image of PDS~70 at the $N\_R$ band, which is also the first high
quality AO-data for $\lambda<1\mu$m taken under good seeing conditions. Combining the archival polarimetric observations across 0.6 to 2.2~$\mu m$, we report the variability in both integrated polarized reflectivity and the azimuthal brightness profiles. We found evidence of significant time variability in the polarized scattered light at optical wavelengths (NR and VBB), while the azimuthal brightness profiles remain stable between the two epochs (Figs 2 and 3). In the near-IR wavelengths (J to Ks), we observed a hint of variability in the H band (H19 vs. H21); however, the current inhomogeneous datasets prevent us from drawing robust conclusions.

Additionally, we found a consistent trend of a brighter near side relative to the far side with wavelength regardless of observing epochs (Fig. 4, top). Interestingly, this near side disappears entirely at visible wavelengths. Furthermore, an excess brightness bump in the southeast ansa in the H band data is absent at other wavelengths (Fig. 4, bottom).

Given the variability of PDS70 and its significant influence on the outer disk scattered light, it is crucial to conduct simultaneous observations across the optical to near-IR wavelengths to extract the disk and dust properties of PDS70. Our results highlight a new complexity involved in analyzing scattered light observations made at different times around variable stars.

\begin{acknowledgement}
We thank the referee for the constructive comments, which have significantly improved the quality and clarity of this letter. We also thank Eric Gaidos for the discussion on photometric variability. 
JM thanks the Swiss National Science Foundation for financial support under grant number P500PT\_222298.
RT and FM acknowledge funding from the European Research Council (ERC) under the European Union's Horizon Europe research and innovation program (grant agreement No. 101053020, project Dust2Planets.)
\end{acknowledgement}

\vspace{-0.8cm}
\bibliographystyle{aa}
\bibliography{biblio}

\appendix
\section{Details on data selection and calibrations}
\subsection{Data selection}
\label{app: data-selection}
\begin{table*}[]
    \centering
    \caption{
    VLT/SPHERE polarimetric imaging data and corresponding observing conditions.}
    \begin{tabular}{l c c l l c c c c }
    \hline
    \hline
    Labels  & Epochs       &  Catg. &   Filters &  $\lambda_c (\Delta \lambda)$[nm]  & nDIT $\times$ DIT & $n_{cyc}$  &  seeing[$\arcsec$] & $\tau_0$[ms] \\
    \hline
    NR &  Mar 01, 2023      & Object*     &     $N\_R$    & 645.9 (56.7)  & 2 $\times$ 32 s      & 11     & 0.40-0.82    & 8.2-18.3 \\
    VBB  &  Jul 09, 2015      & Object*     &        $VBB$  & 735.4 (290.5)    & 6 $\times$ 40 s   & 7\tablefootmark{a}  & 0.81-1.56 & 1.1-2.0 \\
    J16    &    Mar 26, 2016        & Object    &   $J$ + ALC\_YJH\_S  &  1245 (240) & 2 $\times$ 64 s   & 7\tablefootmark{b}    & 1.24-2.21 & 0.9-1.6 \\
           &                        & Flux      &  $J$ + ND\_1.0     & & 3 $\times$ 4 s       & 2      & 1.87-2.39 & 0.9-1.0 \\
    J17     &   Aug 01, 2017         & Object*   &     $J$      & & 20 $\times$ 2 s       & 7     & 0.56-0.77 & 2.0-3.2 \\
    H19    &  Aug 08, 2019        & Object      &  $H$ + ALC\_YJH\_S   & 1625 (290) & 1 $\times$ 64 s   & 9      & 0.39-0.64 & 2.5-3.4 \\
               &                 & Flux         & $H$ + ND\_1.0 &  & 3 $\times$ 2 s        & 1      & 0.62      & 2.5     \\
    H21       & Jul 15, 2021        & Object        & $H$      &  & 1 $\times$ 16 s       & 12     & 0.57-0.80 & 2.5-4.8 \\
                 &                & Object*    & $H$ + ND\_1.0  &  & 1 $\times$ 0.8375 s   & 8\tablefootmark{c}      & 0.57-0.74 & 2.5-3.8  \\
        
    Ks&  Jul 12, 2019      & Object         &  $Ks$+ n/a\tablefootmark{e}    & 2182 (300) & 1 $\times$ 64 s       & 16\tablefootmark{d}     & 0.39-0.72 & 2.8-5.4 \\
               &                 & Flux        & $Ks$+ ND\_1.0  & & 10 $\times$ 2 s       & 2\tablefootmark{d}      & 0.46-0.60 & 3.5-4.6 \\
    \hline
    \end{tabular}
    \tablefoot{Columns give the label name, the observing epoch, the observing type category, the used filters, the central wavelength $\lambda$ and width of the band filter $\Delta \lambda$, the number of detector integration (nDIT) and the detector integration time (DIT), the number of cycles, the seeing, and the coherence time. More specifically, the observing type category includes the flux and object categories. The flux category observes the unsaturated star, without polarimetric cycles. The object category contains 4 polarimetric observations ($Q^+$, $Q^-$, $U^+$, $U^-$) in one cycle. The filter column specifies the used band filter, neutral density filter, and coronagraphs. \\
    \tablefoottext{*}{The object frames are not saturated in the center are used for stellar flux integration. } 
    \tablefoottext{a}{17 cycles were taken but 5 cycles failed in centering. 7 best cycles are selected among the 12 cycles based on the Strehl ratio.}
    \tablefoottext{b}{The last cycle is discarded due to a failure in centering} 
    \tablefoottext{c}{Eight pairs of $Q^{+}$ and $Q^{-}$ measurements, resulting in eight $I_Q$ images.} 
    \tablefoottext{d}{The second part of the observation containing 16 cycles and 1 flux is discarded due to cloudy weather.} 
    \tablefoottext{e}{The observations are executed in polarimetric coronagraphic mode, but the name of the coronagraph is not specified in the header.}
    }
    \label{tab:obs-info}
\end{table*}

Table~\ref{tab:obs-info} summarizes the selected dataset and the corresponding observing conditions. The reasons for discarding some science cycles are given in the footnotes.

\subsection{Polarimetric calibration}
\label{app: polarimetric-calib}
ZIMPOL pipeline does not correct the telescope polarization and IRDAP corrects the instrumental polarization and cross-talks\citep{VanHolstein2020}. For ZIMPOL observations we also estimated and corrected for the telescope polarization which depends on the parallactic angle \cite{Schmid2018}. The interstellar polarization and the intrinsic polarization contribution estimation and removal are optional for both pipelines by normalizing the integrated $Q$ and $U$ in the given region.   
Given that our datasets cover different seeing conditions and include coronagraphic observations, the normalization region is not easy to define uniformly. 
Therefore, we took the non-normalized data and minimized the integrated absolute $U_{\varphi}$ in the disk region to all datasets to correct for the interstellar and intrinsic polarization of the central object. For each cycle of polarimetric observation, we calculated $Q'(x,y) = Q(x,y) - q_{\star}\cdot I_Q(x,y)$, $U'(x,y) = U(x,y) - u_{\star}\cdot I_U(x,y)$ and computed $U'_{\varphi}(x,y)$ with Eq.~\ref{eq:uphi}. We searched for the $q_{\star}$ and $u_{\star}$ to minimize the integrated abs$(U'_{\varphi}(x,y))$ in the disk region. The average and the standard deviation among all cycles for each epoch are summarized in Table~\ref{tab:intrinsic}. 

\begin{figure*}
    \centering
    \includegraphics[width=0.98\textwidth]{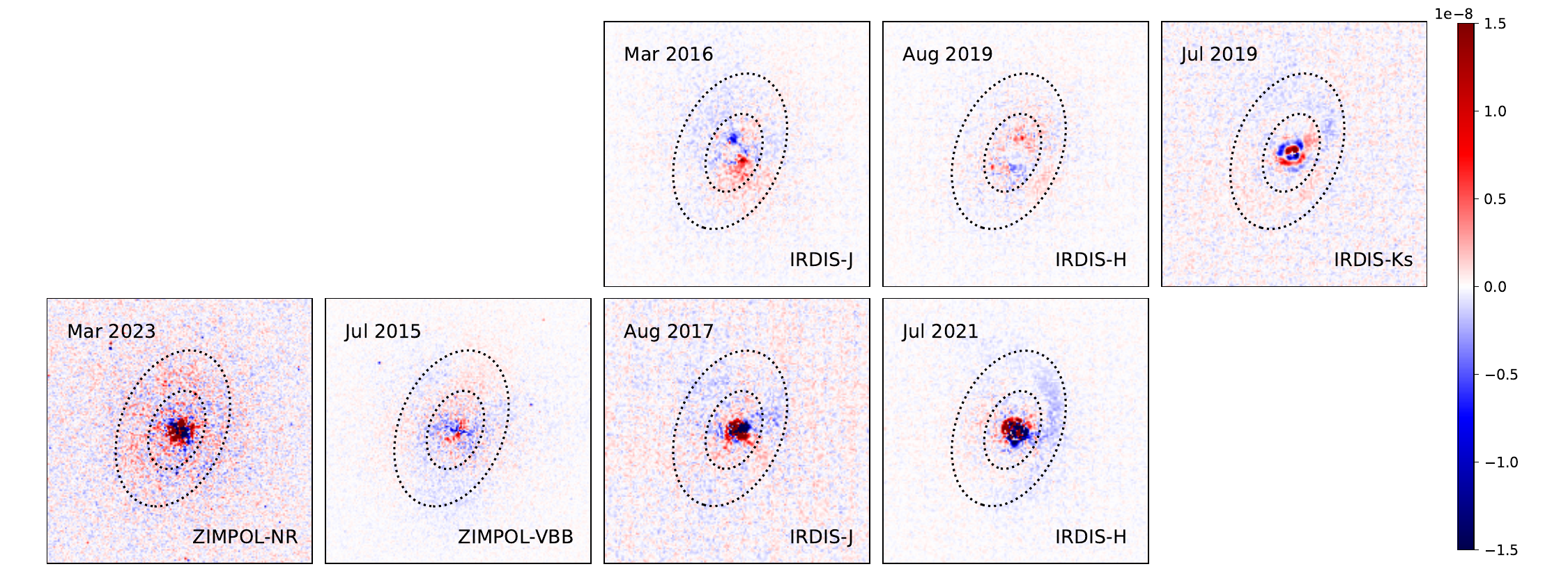}
    \caption{Stokes $U_{\varphi}$ images. All images are normalized in the same way as Fig.~\ref{fig: observations}. The labels and color scales are also the same. The dashed lines, initially used to calculate the azimuthal brightness variation in the H19 panel of Fig.~\ref{fig: observations}, serve to highlight the corresponding area of the disk here.}
    \label{fig: uphi}
\end{figure*}

The mean $U_{\varphi}$ images for each epoch are shown in Fig.~\ref{fig: uphi} corresponding to the 
$Q_{\varphi}$ images in Fig.~\ref{fig: observations}. The same color scale is used to show that the $U_{\varphi}$ signals are generally weak for the outer disk. The rather strong positive and negative $U_{\varphi}$ signal are expected for the inner disk because of PSF convolution effects and the contamination by the intrinsic polarization signal from the unresolved central source. The $U_{\varphi}$ signals have also been reported by \citet{VanHolstein2021} and explained by multiple scattering in the outer disk. However, the expected $U_{\varphi}$ from multiple scattering \citep{Canovas2015} is weaker than what
we observe for the inner disk and what was reported by \citet{VanHolstein2021}.

\begin{figure}
    \centering
    \includegraphics[width=0.98\linewidth]{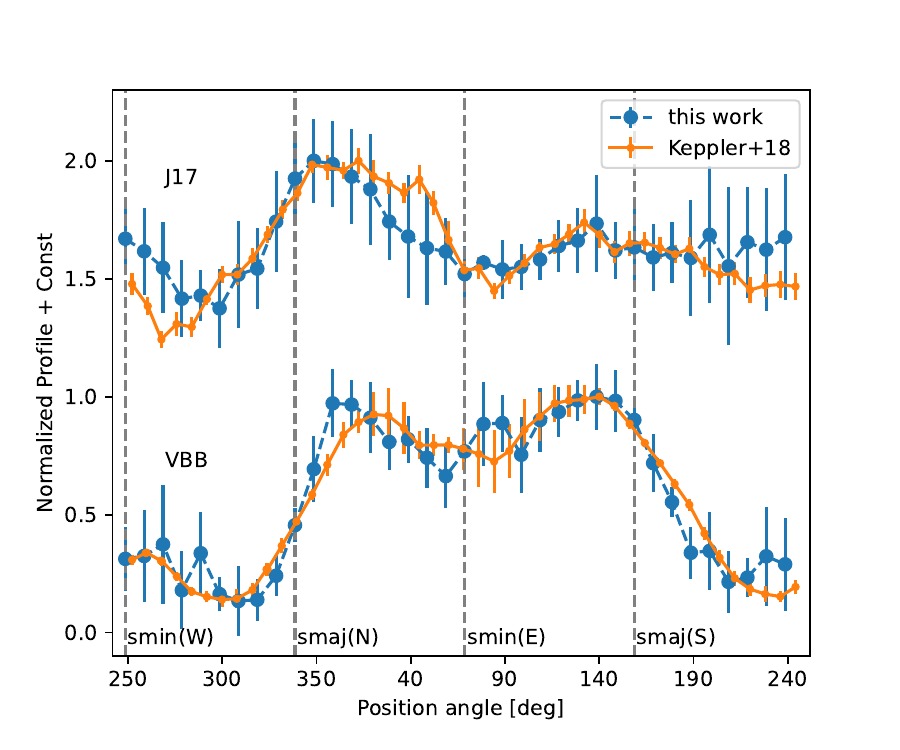}
    \caption{Comparison of azimuthal profiles $Q_{\varphi}(\varphi)$between this work and \cite{Keppler2018}. The J17 profiles are offset by 1. The vertical dashed lines follow the same convention as in Fig.~\ref{fig: variation}}
    \label{fig:compare-profile}
\end{figure}

Our estimations of stellar polarization are in agreement with the previous studies in \citet{Keppler2018, Ma2024}, although those works normalized the data within the central region.
\citet{Pecaut2016} measured low optical extinction for PDS~70, so the $p_{\star}$ is mostly due to intrinsic polarization instead of interstellar polarization. We noticed that $p_{\star}$ for H19 and H21 differs, hinting at variable polarization from the unresolved inner disk. 
Although different normalization step is applied, the effect of data reduction on the azimuthal profile for PDS~70 is minor. To facilitate a clearer comparison, we extracted the curves from Fig 3 in \citet{Keppler2018}, shifted the starting position angle to match our convention, and overplotted our profiles for epoch J17 and VBB in Fig.~\ref{fig:compare-profile} As shown in the figure, the profiles agree with each other at most position angle within our errorbars. We used the standard deviation among all cycles as the uncertainty in our measurements, reflecting the variability observed across cycles. The typical uncertainty is 10-20\% on each point.

\begin{table}[h]
    \centering
    \caption{Intrinsic stellar and interstellar polarization for PDS~70}
    \label{tab:intrinsic}
    \resizebox{0.49\textwidth}{!}{%
    \begin{tabular}{l c c c c}
    \hline
    \hline
    Labels     & $q_{\star}$ [\%]  & $u_{\star}$[\%]  & $p_{\star}$[\%]  & $\theta_{\star}$ [$^\circ$] \\
    \hline
    NR\tablefootmark{a} & $-0.71 \pm 0.05$ & $0.32 \pm 0.07$  & $0.78 \pm 0.05$  & $77.8 \pm 2.7$\\ 
    VBB\tablefootmark{a} & $-0.77 \pm 0.03$ & $0.62 \pm 0.02$  & $0.99 \pm 0.03$  & $70.5 \pm 0.4$\\ 
    J16        & $-0.61 \pm 0.29$ & $0.58 \pm 0.29$  & $0.91 \pm 0.24$  & $68.1 \pm 10.2$ \\
    J17        & $-0.65 \pm 0.23$ & $0.66 \pm 0.29$  & $0.95 \pm 0.31$  & $68.4 \pm 6.8$ \\ 
    H19        & $-0.67 \pm 0.04$ & $0.54 \pm 0.04$  & $0.86 \pm 0.04$  & $70.6 \pm 1.2$  \\
    H21        & $-0.46 \pm 0.09$ & $0.29 \pm 0.09$  & $0.55 \pm 0.10$  & $74.2 \pm 3.5$  \\
    Ks         & $-0.36 \pm 0.66$ & $0.86 \pm 0.63$  & $1.10 \pm 0.71$  & $62.6 \pm 28.3$  \\
    \hline
    \end{tabular}
    }
    \tablefoot{
    \tablefoottext{a}{The ZIMPOL polarization were corrected for the residual telescope polarization $p_{\rm tel} = 0.15\%$ and $\delta_{\rm tel}=52.5^\circ$ for NR and $p_{\rm tel} = 0.36\%$ and $\delta_{\rm tel}=32.9^\circ$ for VBB.}
    }
\end{table}

\subsection{Photometric calibration}
\label{app:photometric-calib}
In this subsection, we detail the integration of stellar flux, $I_{\star}$, the polarized disk flux, $Q_{\varphi}$, and their ratio. 
For each non-saturated total intensity image $I(x,y)$, we estimated and subtracted the background flux, so that the radial profile of the flux falls as a power-law $I(r)\propto r^\alpha$  similar to the PSF halo. The stellar flux $I_{\star}$ is integrated within the 3$\arcsec$ diameter round aperture. The uncertainty is the standard deviation among all frames of a given observing run. 
The $I_{\star}$ for NR, VBB, and J17 are simultaneously obtained with the polarimetric data, allowing monitoring of $I_{\star}$ during the whole observation and showing very stable $\sigma(I_{\star})/\langle I_{\star} \rangle =0.6\%, 1.1\%$ and $0.3\%$, respectively. 
For the H21 data, eight $I_Q$ images show a scatter of $\sigma(I_{\star})/\langle I_{\star}\rangle =5\%$ due to noise at the large separations. No systematic flux increase or decrease with time is seen in the monitoring. The J16 and Ks data have only two flux frames, and as the two measurements agree with each other, we adopt $\sigma(I_\star)/\langle I_{\star}\rangle=3\%$. For H19 which has only one flux frame, we adopt $\sigma(I_\star)/\langle I_{\star}\rangle =5\%$. We note that the $I_{\star}$ value includes disk contribution and, therefore overestimates the stellar intensity. The hot dust emission contributes up to about $10\%$ at K-band \citep{Wang2021}, but less at shorter wavelengths. 

The integrated polarized flux $Q_{\varphi}$ are integrated from the polarized intensity images $Q_{\varphi}(x,y)$ cycle by cycle within the elliptical aperture shown in Fig.~\ref{fig: observations} J16 panel. The uncertainty is the standard deviation of all cycles. Typical uncertainty $\sigma(Q_{\varphi})/\langle Q_{\varphi} \rangle = 7\%$ does not show systematic trends. High uncertainty reaches $15\%$ at $N\_R$ and $J$, mainly due to the low signal-to-noise ratio (S/N). 
For NR, VBB, and J17, simultaneous $Q_{\varphi}$ and $I_{\star}$ measurements are available, the $Q_{\varphi}/I_{\star}$ and the uncertainty are the mean and standard deviation of all cycles. 
For non-simultaneous $I_{\star}$, $Q_{\varphi}/I_{\star}$ is the division of $\langle Q_{\varphi} \rangle$ and $\langle I_{\star} \rangle$-values considering the uncertainty based on error propagation. 

The same process is applied for the corrected integrated polarized flux $\hat{Q}_{\varphi}$. 
The correction follows the method described in \citet{Ma2024}. The method does not require detailed parametric modeling, but only (quasi-)~simultaneous PSFs to correct the smearing effect on the integrated flux. The VBB band is the most affected by the PSF smearing due to the poor seeing condition and the observed polarized flux is only 44\% of the intrinsic polarized flux (see Table~\ref{tab:aperture-count}). When applying A-type correction, $\hat{Q}_{\varphi}/I_{\star}$ is shown to have a smaller uncertainty than $Q_{\varphi}/I_{\star}$ for NR and VBB because simultaneous PSF corrects the smearing effect better. For non-simultaneous $I_{\star}$ and disk polarimetry data the smearing
correction only retrieves the lost flux but does not improve the accuracy (B-type correction in \cite{Ma2024}). All the integrated fluxes are summarized in Table~\ref{tab:aperture-count}. 

\begin{table*}[]
    \caption{Aperture count rates for the total intensity, $I_{\star}$, the azimuthal polarization, $Q_{\varphi}$, and the ratio, $Q_{\varphi}/I_{\star}$, for PDS~70 for each observing date.}
    \label{tab:aperture-count}
    \centering
    \begin{tabular}{c c c c c c c}
    \hline
    \hline
    Label    
        & $I_{\star}$ [$\times 10^4$ $s^{-1}$] 
            & $Q_{\varphi}$ [$\times 10^4$ $s^{-1}$]    
                & $Q_{\varphi}/I_{\star}$ [\%] 
                    & $ \hat{Q}_{\varphi}$ [$\times 10^4$ $s^{-1}$] 
                        & $\hat{Q}_{\varphi}/I_{\star}$ [\%] 
                            & Proc \tablefootmark{a} \\                    
    \hline
    NR    & 30.6 $\pm$ 0.2   & 0.029 $\pm$ 0.004 & 0.093 $\pm$ 0.012 & 0.050 $\pm$ 0.005  & 0.163 $\pm$ 0.016   & A\\ 
    VBB     & 124.6 $\pm$ 1.4  & 0.214 $\pm$ 0.016  & 0.171 $\pm$ 0.012 & 0.483 $\pm$ 0.006  & 0.388 $\pm$ 0.005  & A\\ 
    J16       & 80.0 $\pm$ 2.4   & 0.139 $\pm$ 0.023  & 0.174 $\pm$ 0.034 & 0.257 $\pm$ 0.044  & 0.321 $\pm$ 0.065  & B2\\ 
    J17      & 175.8 $\pm$ 0.6  & 0.336 $\pm$ 0.026 & 0.191 $\pm$ 0.015 & 0.585 $\pm$ 0.055  & 0.333 $\pm$ 0.031   & A\\ 
    H19       & 111.3 $\pm$ 5.6  & 0.187 $\pm$ 0.014 & 0.168 $\pm$ 0.021 & 0.302 $\pm$ 0.022  & 0.271 $\pm$ 0.033   & B1\\ 
    H21      & 234.8 $\pm$ 10.7 & 0.611 $\pm$ 0.020 & 0.260 $\pm$ 0.020 & 0.824 $\pm$ 0.028  & 0.355 $\pm$ 0.028   & B2 \\
    Ks     & 73.6 $\pm$ 2.2   & 0.220 $\pm$ 0.017 & 0.299 $\pm$ 0.032 & 0.295 $\pm$ 0.023  & 0.401 $\pm$ 0.043   & B2\\ 
    \hline
    \end{tabular}
    \tablefoot{
    \tablefoottext{a}{Correction method applied to the integrated polarized flux as described in \cite{Ma2024}. }
    }
\end{table*}

\section{Correction for azimuthal profile}
\label{app: corr-azi}
The correction method developed by \cite{Ma2024} corrects for the smearing effect on the integrated flux but doesn't address the smearing along the azimuthal direction. To investigate the PSF smearing effects on the azimuthal signal distribution we need a model fitting procedure similar to \citet{Ma2023} which takes the intrinsic azimuthal dependence of the polarization into account. We simulate the asymmetric asymmetric brightness along the disk with analytical formulation $A(\varphi)$:
\begin{equation}
    A(\varphi) = A_0 + A_1\cdot \exp\left(-\frac{(\varphi - \varphi_1)^2}{2\sigma_1^2}\right) + A_2\cdot \exp\left(-\frac{(\varphi - \varphi_2)^2}{2\sigma_2^2}\right) \, ,
\end{equation}
where $\varphi$ is the position angle measured from north to east, 
with $A_1$, $A_2$, $\varphi_1$, $\varphi_2$, $\sigma_1$, and $\sigma_2$ representing the intensities, azimuthal positions, and widths of two brightness peaks, respectively, and $A_0$ indicating the baseline brightness level. 

In the radial direction, the disk is characterized by a double power law function: 
\begin{equation}
    Q_{\varphi}/I_{\star}(r,\varphi) = A(\varphi)\cdot \left( \left( \dfrac{r}{r_0}\right)^{-2\alpha_{\rm in}} + \left( \dfrac{r}{r_0}\right)^{-2\alpha_{\rm out}} \right)^{-1/2} \,.
\end{equation}
The $r$ value is the deprojected radius using diskmap, $r_0$ is the approximate peak radius, and the two power-law indices $\alpha_{\rm in} >0$ and $\alpha_{\rm out}<0$ define the inner and outer disk slope, respectively.

The inner edge is not spatially resolved and we fixed $\alpha_{in}$= 15 for all wavelengths. The remaining seven parameters $r_0, \alpha_{out}, A_0, A_1, A_2, \varphi_1, and \varphi_2$ are variable. The analytical model images are convolved with their corresponding PSFs and fitted to the averaged $Q_{\varphi}/I_{\star} (x,y)$ image for all selected epochs in the deprojected 35-70 au aperture as shown in Fig.~\ref{fig: observations}. The fitted parameters are summarized in Table~\ref{tab: analytical}. We note that the model aims not for a perfect azimuthal profile match, but to assess the relative change of the azimuthal profile caused by PSF convolution. 

\begin{figure}
    \centering
    \includegraphics[width=0.48\textwidth]{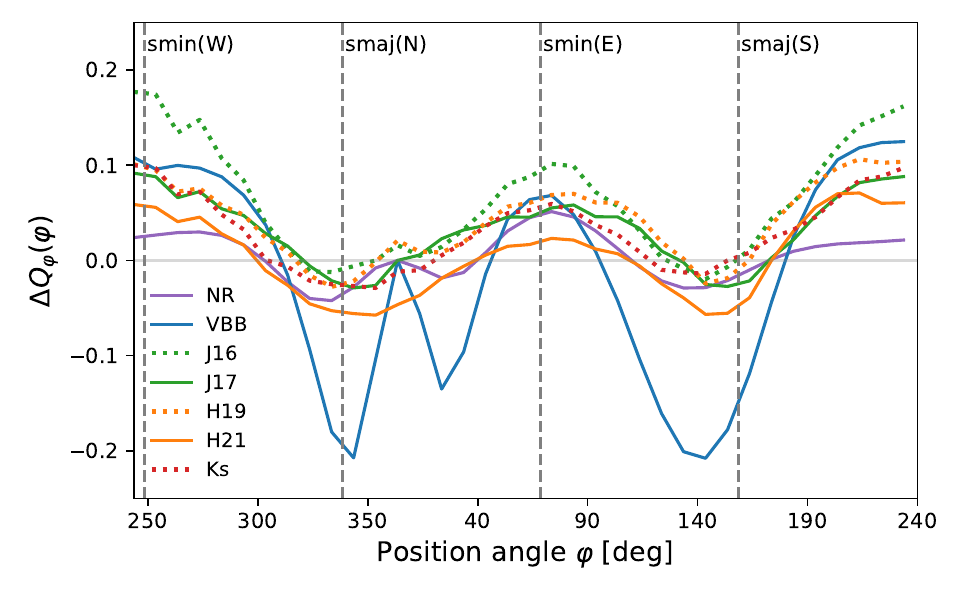}
    \caption{Relative difference, $\Delta Q_{\varphi}(\varphi) = \hat{Q}_{\varphi}(\varphi) - \Tilde{Q}_{\varphi}(\varphi),$ between the intrinsic azimuthal profile and the convolved azimuthal profile. }
    \label{fig: correct-curve}
\end{figure}

To assess the relative change, we convolved the best-fit analytical model with the corresponding PSFs and obtained the convolved model images $\Tilde{Q}_{\varphi}(x,y)$. We follow the same procedure to calculate the convolved azimuthal profile $\Tilde{Q}_{\varphi}(\varphi)$, depicted in Fig.~\ref{fig: variation} as lines. The profiles $\Tilde{Q}_{\varphi}(\varphi)$ and $\Tilde{Q}_{\varphi}(\varphi)$ are normalized to their peak values. The relative difference $\Delta Q_{\varphi}(\varphi) = \hat{Q}_{\varphi}(\varphi) - \Tilde{Q}_{\varphi}(\varphi)$ illustrates the effect of PSF smearing on the azimuthal profile. Figure~\ref{fig: correct-curve} displays $\Delta Q_{\varphi}(\varphi)$ for each epoch, indicating a trend where brightness along the major axis gets weaker and that along the minor axis gets stronger, consistent with the "smearing" concept. The near side loses slightly more flux than the far side due to the inclined disk geometry, which makes the near side appear closer to the center. In general, the PSF smearing alters brightness by less than $10\%$ across all position angles. Exceptions are the $VBB$ and $J$ band, where the relative difference reaches $20\%$ near the major and minor axis respectively. Thus, the PSF smearing effect on PDS~70 is not substantial and preserves the overall profile shape before and after the convolution. 

To correct the smearing effect on the azimuthal profiles, we multiplied the factor $c(\varphi) = \hat{Q}_{\varphi}(\varphi)/\Tilde{Q}_{\varphi}(\varphi)$ with the observed $Q_{\varphi}(\varphi)$. Then we measured the brightness contrast as described in Sect.~\ref{section: azimuthal}. The corrected brightness contrasts are summarized in Table~\ref{tab: contrast}. Here, $\hat{\alpha}_{S-N}$ doesn't change much after correction, while $\hat{\alpha}_{W-E}$ is slightly higher after the correction because the near-side (W) is closer to the center and therefore loses more flux. 

\begin{table*}[]
    \centering
    \caption{Best fitting parameters for analytical model. }
    \label{tab: analytical}
    \begin{tabular}{c c c c c c c c c c}
    \hline
    \hline
    Label  &  $A_0$ & $A_1$ & $\varphi_1$ & $\sigma_1$ & $A_2$ & $\varphi_2$ & $\sigma_2$ & $r_0$ & $\alpha_{\rm out}$ \\
            & $10^{-7}$mas$^{-2}$ & $10^{-7}$mas$^{-2}$ & deg & deg & $10^{-7}$mas$^{-2}$ & deg & deg & AU & - \\
    \hline
    NR  & 0.1 & 2.3 & 3.1 & 20.3 & 2.1 & 97.2 & 41.5 & 49.5 & -7.8  \\
    VBB                    & 0.9 & 2.9 & 9.0 & 12.2 & 3.3 & 105.2 & 61.4 & 49.7 & -7.1  \\
    J16                      & 16.8 & 10.0 & 357.4 & 19.2 & 0\tablefootmark{a}  & -     & -    & 48.8 & -4.6  \\
    J17                     & 8.8 & 18.6 & 7.0 & 31.7 & 15.2 & 130.5 & 28.4 & 49.2 & -4.7 \\
    H19                      & 7.2 & 5.4 & 5.1 & 15.9 & 12.5 & 143.1 & 33.8 & 49.7 & -3.3 \\
    H21                     & 11.3 & 3.6 & 348.0 & 16.0 & 11.1 & 180.9 & 27.5 & 49.6 & -2.8 \\
    Ks                     & 14.9 & 8.4 & 350.8 & 20.6 & 9.8 & 156.1 & 17.2 & 50.4 & -3.5 \\
    \hline
    \end{tabular}
    \tablefoot{
    \tablefoottext{a}{Only one brightness peak is needed to fit J16 and $A_2$ is set to 0 in the fitting process. }
    }
\end{table*}

\begin{table}[h]
    \centering
    \caption{Brightness contrast.}
    \label{tab: contrast}
    \resizebox{0.49\textwidth}{!}{%
    \begin{tabular}{l c c c c}
    \hline
    \hline
    Label     & $\alpha_{W-E}$  & $\alpha_{S-N}$  &  $\hat{\alpha}_{W-E}$   & $\hat{\alpha}_{S-N}$ \\
    \hline
    NR & <-0.66\tablefootmark{a} & $-0.17 \pm 0.21$ & <-0.45\tablefootmark{a} & $-0.15 \pm 0.21$ \\
    VBB  & $-0.45 \pm 0.17$ & $-0.06 \pm 0.06$ & $-0.18 \pm 0.22$ & $-0.06 \pm 0.05$ \\
    J16  & $0.11 \pm 0.13$ & $-0.18 \pm 0.08$ & $0.19 \pm 0.13$ & $-0.18 \pm 0.08$ \\
    J17  & $-0.11 \pm 0.25$ & $-0.11 \pm 0.08$ & $0.00 \pm 0.26$ & $-0.10 \pm 0.08$\\ 
    H19  & $0.11 \pm 0.11$ & $0.25 \pm 0.04$ & $0.19 \pm 0.10$ & $0.24 \pm 0.04$ \\
    H21  & $0.13 \pm 0.21$ & $0.21 \pm 0.05$ & $0.17 \pm 0.21$ & $0.25 \pm 0.05$ \\
    Ks   & $0.17 \pm 0.14$ & $-0.02 \pm 0.05$ & $0.22 \pm 0.13$ & $-0.01 \pm 0.05$ \\
    \hline
    \end{tabular}}
    \tablefoot{\tablefoottext{a}{Upper limit given}}
\end{table}

\section{Flat disk modeling}
\label{app: flat-disk}
This section aims to present flat surface disk results based on the model grid calculation from \cite{Ma2022}, which could qualitatively reproduce the observed flux and near-to-far contrast as a function of wavelength. 

\begin{figure}
    \centering
    \includegraphics[width=0.49\textwidth]{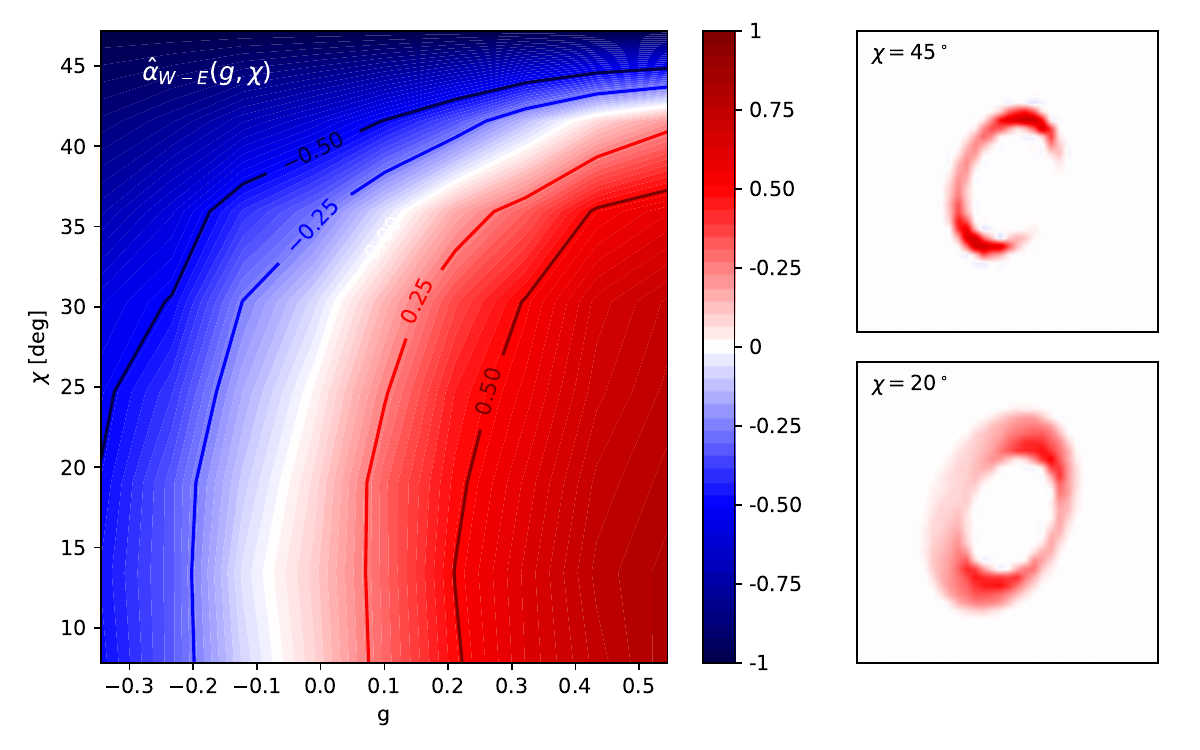}
    \caption{Modelling result assuming plane-parallel surface disk. Left:  Intrinsic brightness contrast $\hat{\alpha}_{W-E}$ as function of the asymmetric parameter $g$ and the disk wall slope $\chi$. Levels are indicated with solid lines with values. Right: Examples of simulated polarized disk images, with parameters $g=0.1$ and $\chi = 45^\circ$ in the upper panel and $g=0.1$ and $\chi=20^\circ$ in the lower panel. }
    \label{fig: flatmodel}
\end{figure}

For the disk geometry, we fixed the inclination to $i=47.5^\circ$, which is the closest value that the plane-parallel model can accommodate to the $i=49.7^\circ$ observed in reflected disk images. We also fixed the disk opening angle $\alpha=\arctan(h/r)=8^\circ$, which follows from the estimated $h=7$au and $r=50$au derived by ellipse fitting \citep{Takami2014, Keppler2018}. For the scattering parameters, we fixed the single scattering albedo $\omega=0.5$ and the maximum polarization $p_{\rm max} = 0.5$, as they do not greatly affect the brightness contrast \citep{Ma2022}. In the model grid, we varied the asymmetry parameter, $g\in [-0.2, 0.5]$, with a step of 0.1 and disk wall slope of $\chi \in [10^\circ, 50^\circ],$ with a step of $5^\circ$, creating a coarse grid of disk models. For each model, we calculated the brightness contrast $\hat{\alpha}_{W-E}$, following the same process described in Sect.~\ref{sect: analysis}, and the results $\hat{\alpha}_{W-E} (g, \chi)$ are plotted in Fig.~\ref{fig: flatmodel}.  

To reproduce $\hat{\alpha}_{W-E} < -0.25$ seen in the visible, backward scattering dust at the level $g<-0.2$ is needed for a flat disk wall $\chi\approx10^\circ$; otherwise, a high disk wall $\chi>35^\circ$ is required for small dust $g\approx 0$. Even higher $\chi$ is needed for forward scattering dust. Figure~\ref{fig: flatmodel} presents two examples of flat models, which could resemble the observations. It should be noted that these simple disk models assume a plane-parallel disk surface structure. Therefore, the simulations produce no signal from the disk near side if $\chi > 90-i = 40^\circ$ for PDS 70. For a reliable estimate of the dust asymmetry parameter of a moderately inclined disk like PDS70, we should consider more detailed models with curved disk surface geometry and a vertical stratification for the scattering dust particles.

%
%

\end{document}